\documentclass[preprint,12pt]{elsarticle}

\usepackage{amssymb}
\usepackage{amsmath}

\usepackage{lineno}

\usepackage{caption}
\journal{Nuclear Instruments and Methods in Physics Research A}

\begin{document}

\begin{frontmatter}



\title{Analysis of the Origin of the Peak Shift under Back-side Illumination in the SOI Pixel Detector XRPIX}


\author[inst1]{Fumiya Shiga\corref{cor1}} 
\ead{6224515@alumni.tus.ac.jp}
\cortext[cor1]{Corresponding author.}
\author[inst1,inst5]{Takayoshi Kohmura}
\author[inst2,inst5]{Yuusuke Uchida}
\author[inst3]{Ikuo Kurachi}
\author[inst4,inst5]{Kouichi Hagino}

\affiliation[inst1]{organization={Department of Physics and Astronomy, Faculty of Science and Technology, Tokyo University of Science},
            addressline={2641 Yamazaki}, 
            city={Noda},
            postcode={278-8510}, 
            state={Chiba},
            country={Japan}}
\affiliation[inst2]{organization={Institute of Space and Astronautical Science (ISAS) / Japan Aerospace Exploration Agency (JAXA)},
            addressline={3-1-1 Yoshinodai, Chuo-ku}, 
            city={Sagamihara},
            postcode={252-5210}, 
            state={Kanagawa},
            country={Japan}}
\affiliation[inst3]{organization={D\&S Inc.},
            addressline={774-3-213 Higashiasakawacho}, 
            city={Hachioji},
            postcode={193-0834}, 
            state={Tokyo},
            country={Japan}}
\affiliation[inst4]{organization={RIKEN Pioneering Research Institute},
            addressline={2-1 Hirosawa}, 
            city={Wako},
            postcode={351-0198}, 
            state={Saitama},
            country={Japan}}
\affiliation[inst5]{organization={Research Center for Space System Innovation, Research Institute for Science \& Technology, Tokyo University of Science},
            addressline={2641 Yamazaki}, 
            city={Noda},
            postcode={278-8510}, 
            state={Chiba},
            country={Japan}}

\begin{abstract}
X-ray spectra obtained by Front Illumination (FI) or Back Illumination (BI) configurations have been deeply studied by using full depletion of around $300~\mathrm{\mu m}$ silicon-on-insulator (SOI) pixel sensor called as XRPIX8 which has high detection efficiency over a wide energy range from soft to hard X-rays. The results show that the X-ray peak positions in the BI configuration shift toward lower energies, with a larger shift for lower energy X-rays. To investigate the origin of this peak shift, we evaluated the charge collection efficiency based on the trap density distribution obtained from reverse-bias I-V measurements and considered the effect of charge cloud spreading based on the electric field distribution. The results indicate that the contribution of charge loss due to back-side defects is limited and that the reduction in collected charge due to charge sharing associated with charge cloud spreading is the primary cause of the peak shift. These results are useful for high precision energy calibration and detector response modeling of BI-type XRPIX in future X-ray astronomical missions.
\end{abstract}


\end{frontmatter}



\section{Introduction}
In X-ray astronomy, imaging and spectroscopic analyses of X-rays from astronomical objects are used to investigate their physical states and elemental compositions. Because continuum emission from different physical processes and characteristic emission lines from different elements span a broad energy range from a few hundred eV to several tens of keV, observations over a broad energy band are required. Therefore, X-ray imaging sensors require high detection efficiency and excellent spectroscopic performance over a wide energy range.
A Back-Illumination (BI) configuration, in which X-rays are incident from the side opposite to the circuit layer, is advantageous for detecting low-energy X-rays because it reduces X-ray absorption in the circuit layer compared with a Front-Illumination (FI) configuration. However, a thick depletion layer is required for efficient detection of high-energy X-rays. In conventional BI-type Charge Coupled Devices (CCDs), the depletion layer thickness is limited by the device structure, resulting in lower detection efficiency for high-energy X-rays compared with FI-type CCDs~\cite{Koyama}.

The X-Ray SOI PIXel sensor (XRPIX)~\cite{Arai,Tsuru}, based on Silicon-On-Insulator (SOI) technology, overcomes this limitation by enabling a sufficiently thick depletion layer even in the BI configuration. This is because the sensor layer and the silicon layer containing the readout circuit are separated by an insulating layer, allowing the substrate concentration of the sensor layer to be independently controlled without affecting the circuit characteristics. In fact, XRPIX has achieved full depletion of the entire sensor thickness of $300~\mathrm{\mu m}$~\cite{Kodama,Yukumoto2024}.

To investigate the physical states of astrophysical objects through X-ray spectroscopy, an accurate understanding of the X-ray response of the sensor is essential for accurately deriving information about the incident X-rays from the observed sensor data. Thus, to understand the X-ray response of the BI configuration in XRPIX, we compared the spectra obtained with the FI and BI configurations. We found that the X-ray peak positions in the BI configuration were shifted toward lower energies compared with those in the FI configuration. In this study, we investigate the origin of this peak shift.

\section{Device structure and experimental procedure}
\begin{figure}[tbh]
    \centering
    \includegraphics[width=0.7\linewidth]{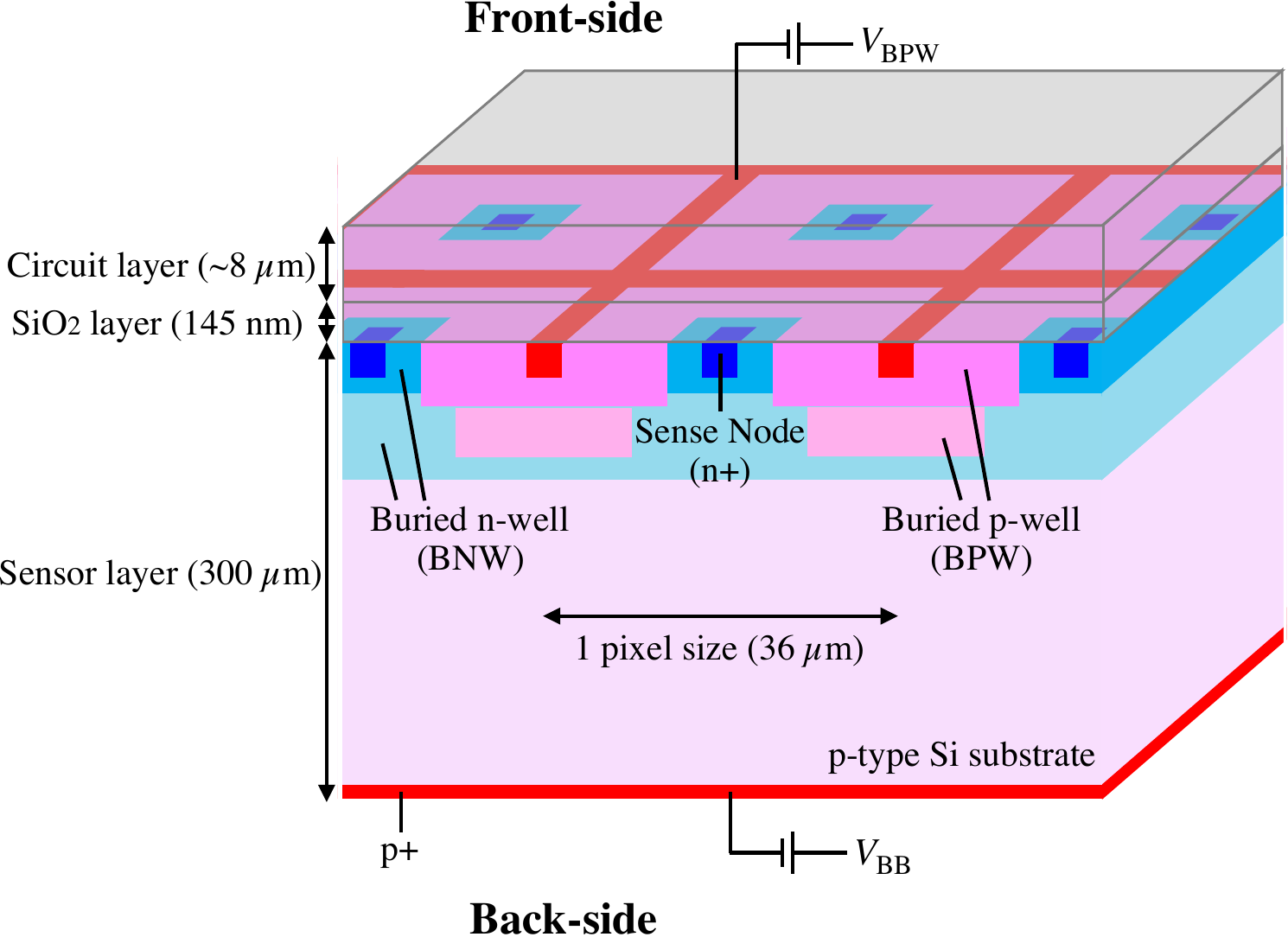}
    \caption{Cross-sections of XRPIX8.}
    \label{fig:01}
\end{figure}

A prototype device, XRPIX8\footnote{The specific version used in this study is 
XRPIX8.5.}, which features the optimized pixel design of the XRPIX series was used in this experiment \cite{Yukumoto2024}. Figure~\ref{fig:01} shows a schematic cross-sectional view of the XRPIX8 which is an X-ray detector based on Silicon-On-Insulator (SOI) technology and has a monolithic structure consisting of a Si circuit layer, a $\mathrm{SiO_2}$ insulating layer between the Si layers, and a Si sensor layer. The pixel size of XRPIX8 is $36~\mathrm{\mu m}\times36~\mathrm{\mu m}$, and the effective number of pixels is 80$\times$94. 
The sensor layer thickness is $300~\mathrm{\mu m}$ and fabricated by high resistivity (approximately $10~\mathrm{k\Omega\cdot cm}$) p-type silicon grown by the Floating Zone method. Because of the high-resistivity Si, it was experimentally confirmed that the fully depleted sensor is obtained with a back bias voltage of approximately $250~\mathrm{V}$. The sensor also employs a Pinned Depleted Diode (PDD) structure \cite{Kamehama}, which suppresses the back-gate effect, reduces dark current, and achieves high charge-collection efficiency \cite{Kodama,Yukumoto2024,Harada,Yukumoto2025}. On the back-side, a boron-doped $\mathrm{p^+}$ diffusion layer with a thickness of approximately $0.25~\mathrm{\mu m}$ and a $200$-$\mathrm{nm}$-thick Al layer for visible-light shielding are formed. Since the front-side contains five metal interconnect layers, an $8$-$\mathrm{\mu m}$-thick silicon oxide film, including the surface passivation layer, is formed.

\begin{figure}[tbh]
    \centering
    \includegraphics[width=0.7\linewidth]{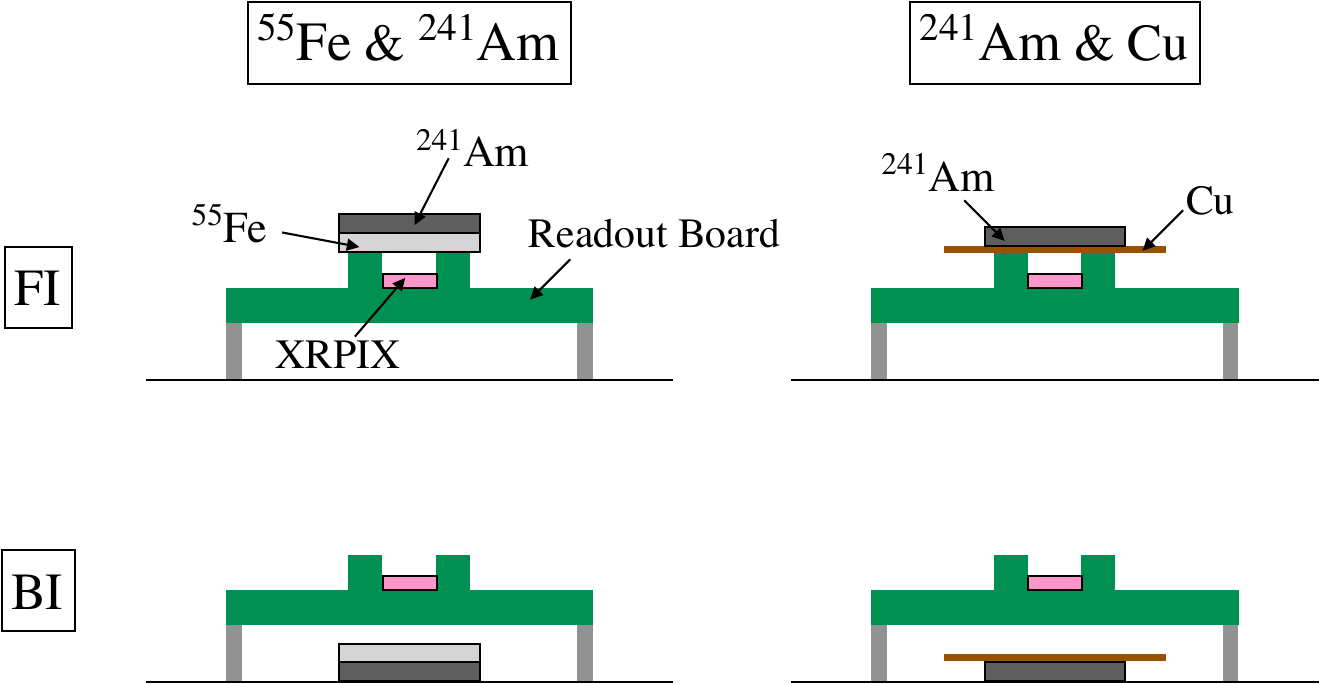}
    \caption{A schematic view of the experimental setup for FI and BI measurements.}
    \label{fig:02}
\end{figure}

X-ray energy spectra were measured using the X-ray sources listed in Table~\ref{tab:xray_energy}. Figure~\ref{fig:02} shows the experimental setup for the FI and BI measurements. The irradiated X-rays were generated using ${}^{55}\mathrm{Fe}$ and ${}^{241}\mathrm{Am}$ radioactive sources. In addition, a 0.5-mm-thick Cu plate was inserted between the ${}^{241}\mathrm{Am}$ source and XRPIX8, and characteristic Cu X-rays generated by photoelectric absorption of ${}^{241}\mathrm{Am}$ X-rays in the Cu plate were also used as the X-ray source.
The XRPIX8 mounted on the readout board was placed in a thermostatic chamber and cooled to $-15.0~{}^\circ\mathrm{C}$. A back bias voltage of $-300~\mathrm{V}$ was applied to the photodiode to achieve full depletion of the sensor layer. The exposure time for X-ray spectrum acquisition was set to $1000~\mu\mathrm{s}$. As shown in Fig.~\ref{fig:02}, FI and BI measurements were performed by changing only the side on which the X-ray source was placed, while keeping all other experimental conditions unchanged.
For the digital values (pulse heights) read out from each pixel, corresponding to the charge generated by X-ray absorption, events whose pulse heights exceeded a predefined event threshold were selected as X-ray events. Furthermore, if the pulse heights of the surrounding neighboring pixels were below the split threshold, the event was classified as a single-pixel event.

\begin{table}[tbh]
    \centering
    \caption{Energies of the irradiated X-rays.}
    \label{tab:xray_energy}
    \begin{tabular}{ccc}
    \hline
    Source&Emission Line&Energy [keV]\\
    \hline
    $^{55}$Fe&Mn-K$\alpha$&5.89\\
    Cu&Cu-K$\alpha$&8.04\\
    $^{241}$Am&Np-L$\alpha$&13.94\\
    $^{241}$Am&Np-L$\beta$&17.75\\
    \hline
    \end{tabular}
\end{table}

\section{Results}
\begin{figure}[tbh]
    \centering
    \includegraphics[width=0.7\linewidth]{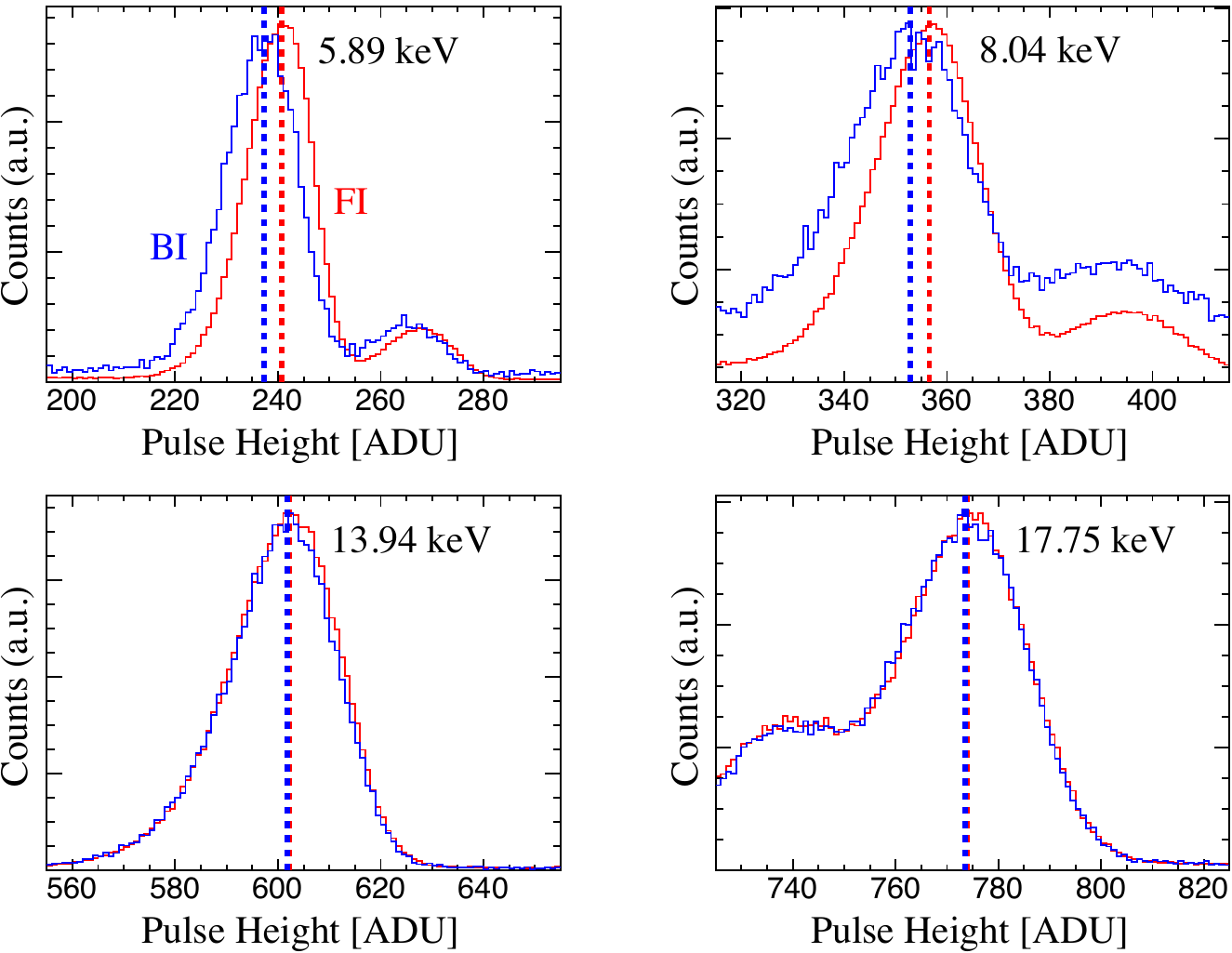}
    \caption{Single-pixel event spectra for each X-ray energy. The red and blue spectra correspond to FI and BI, respectively. The counts are in arbitrary units.}
    \label{fig:03}
\end{figure}

Figure~\ref{fig:03} shows the X-ray spectra acquired under FI and BI conditions. The spectra were constructed using only single-pixel events. As shown in the figure, noticeable differences are observed between the FI and BI spectra. In the BI case, the peak positions are shifted toward lower energies.

To quantitatively evaluate the peak shift toward lower energies, an evaluation parameter, shift ratio, is introduced. First, each characteristic X-ray peak was fitted with a Gaussian function, and the peak position was determined from the fitted mean value. For each X-ray energy, the shift ratio was defined as
\begin{equation}
\mathrm{Shift\ Ratio} = \frac{\mathrm{Mean}~\mathrm{BI} - \mathrm{Mean}~\mathrm{FI}}{\mathrm{Mean}~\mathrm{FI}},
\end{equation}
where $\mathrm{Mean}~\mathrm{FI}$ and $\mathrm{Mean}~\mathrm{BI}$ represent the mean values obtained under the FI and BI conditions, respectively.

\begin{figure}[tbh]
    \centering
    \includegraphics[width=0.5\linewidth]{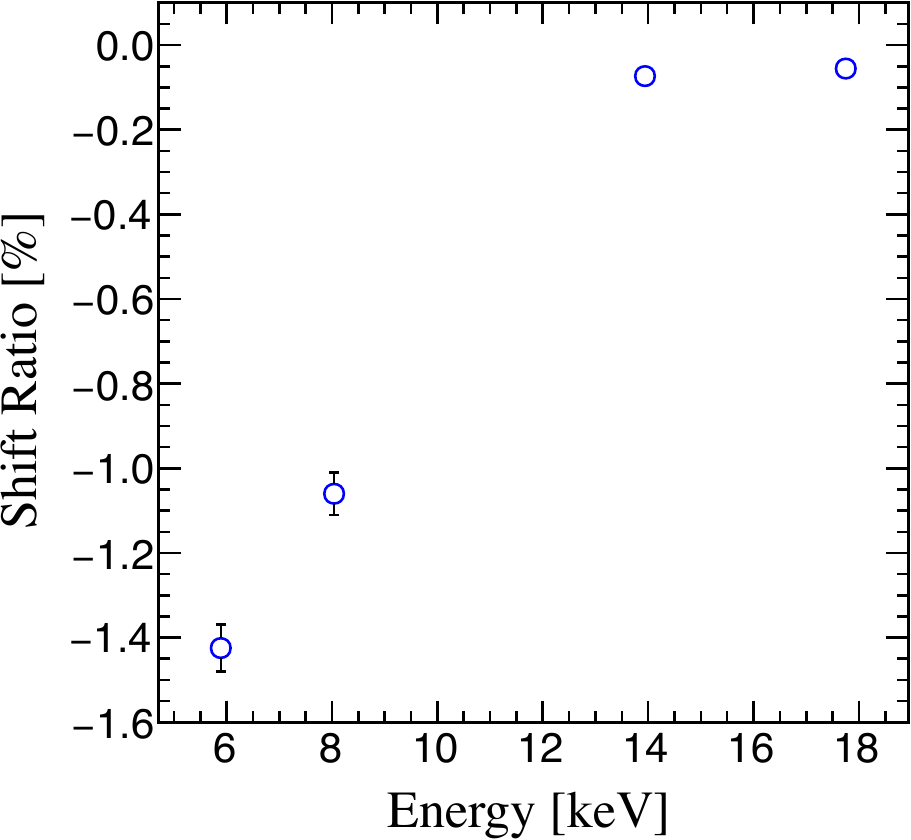}
    \caption{Dependence of the shift ratio on X-ray energy.}
    \label{fig:04}
\end{figure}

The resulting shift ratios are shown in Fig.~\ref{fig:04}. As illustrated in the figure, the shift ratio increases as the X-ray energy decreases. At the lowest measured energy of 5.89~keV, the shift ratio reached approximately 1.4\%.
The energy dependence of the peak shift ratio may complicate the construction of the energy calibration function and the detector response model. Understanding the physical origin of this peak shift is therefore important for accurate energy calibration and reliable modeling of the X-ray response in the BI configuration. In this study, we investigate the origin of the peak shift observed for low-energy X-rays in the BI configuration.

\section{Discussion}
It was confirmed that the shift ratio increased as the X-ray energy decreased. This energy dependence suggests that the peak shift is related to the X-ray absorption depth. In the BI configuration, lower-energy X-rays are more likely to be absorbed near the back-side. This implies that the peaks formed by charges generated closer to the back-side are shifted to lower energies compared with those obtained in the FI.

This peak shift is considered to originate either from 
charge loss during transport or from incomplete charge collection at the sense node. In the former case, the generated charge may be lost through carrier recombination at defects introduced during the back-side fabrication process. In the latter case, the charge cloud generated near the back surface may spread while drifting over a distance comparable to the sensor thickness. In the following, the mechanisms responsible for the peak shift in the BI are quantitatively investigated from these two viewpoints.

\subsection{Effect of back-side defect}
A heavily doped $\mathrm{p^+}$ layer was formed on the back-side of the sensor to suppress the increase in dark current caused by the depletion region reaching the back-side Si surface under full depletion, while simultaneously serving as the back-side electrode. To form this layer, the wafer was first grind to its final thickness, followed by high-dose boron ion implantation and laser annealing for dopant activation. The laser annealing was performed by scanning a laser beam across the back-side of the wafer. Consequently, crystal defects introduced during the grinding and ion implantation processes may not have been fully recovered. In addition, residual metallic contamination remaining on the back-side after grinding may diffuse into the Si during laser annealing, creating trap levels in the Si bandgap. These crystal defects and trap levels can act as recombination centers for X-ray-generated charges, resulting in charge loss for events generated near the back-side.

To evaluate the depth profile of trap density, reverse I-V measurements were carried out using a PN-junction diode fabricated in the sensor. The diode was designed to suppress surface generation current originating from the lateral expansion of the depletion region along the Si surface, allowing only the bulk generation-recombination current to be measured. The depth profile of trap density was derived from the rate of increase of the reverse current as the depletion region was extended into the sensor by stepwise increases in the back bias voltage. The measurements were performed at $27.0~{}^\circ\mathrm{C}$ while the back bias was swept from 0 to -300~V in 5-V increments. Assuming that the reverse current $I$ is a function of the depletion width $W$, the depth profile of the trap density $N_\mathrm{t}(z)$ was calculated using
\begin{equation}
	I(W)=q n_i \sigma_\mathrm{t} v_\mathrm{th} \int_0^WN_\mathrm{t}(z)dz,
    \label{eq:Nt}
\end{equation}
where $q$, $n_i$, $\sigma_\mathrm{t}$, and $v_\mathrm{th}$ are the elementary electron charge, intrinsic carrier concentration, trap capture cross section, and electron thermal velocity, respectively. The values of $n_i=9.65\times10^{9}~\mathrm{cm^{-3}}$, $\sigma_\mathrm{t}=1.0\times10^{-15}~\mathrm{cm^2}$, and $v_\mathrm{th}=2.28\times10^7~\mathrm{cm/s}$ were used in the calculation~\cite{Sze,Grove}.
\begin{figure}[tbh]
    \centering
    \includegraphics[width=0.5\linewidth]{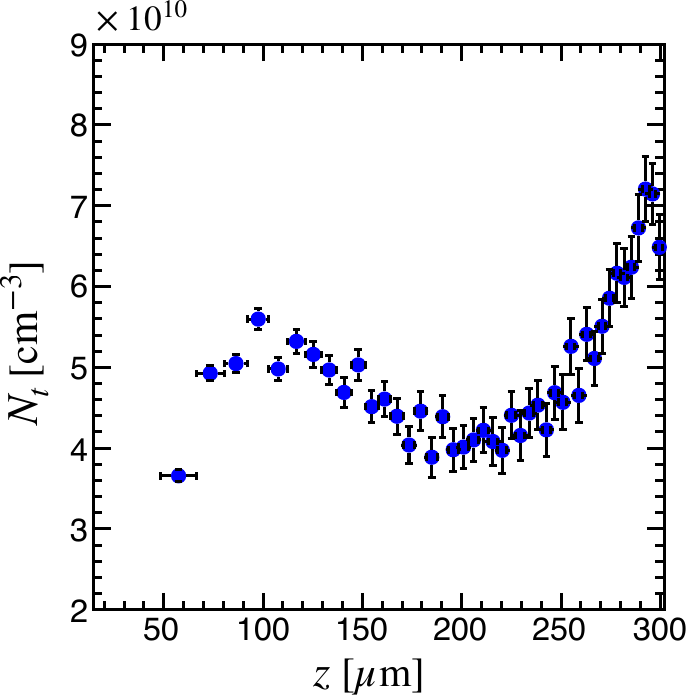}
    \caption{Depth profile of the trap density $N_\mathrm{t}(z)$.}
    \label{fig:05}
\end{figure}

As shown in Fig.~\ref{fig:05}, a region with high trap density was observed near the back-side ($z\sim300~\mu\mathrm{m}$).
To evaluate the impact of this region on charge collection, the depth-dependent charge collection efficiency (CCE) was calculated using the measured trap density profile $N_\mathrm{t}(z)$. The charge transport calculation was implemented based on the method used in the detector response simulator ComptonSoft \cite{Odaka}.

\begin{figure}[tbh]
    \centering
    \includegraphics[width=0.55\linewidth]{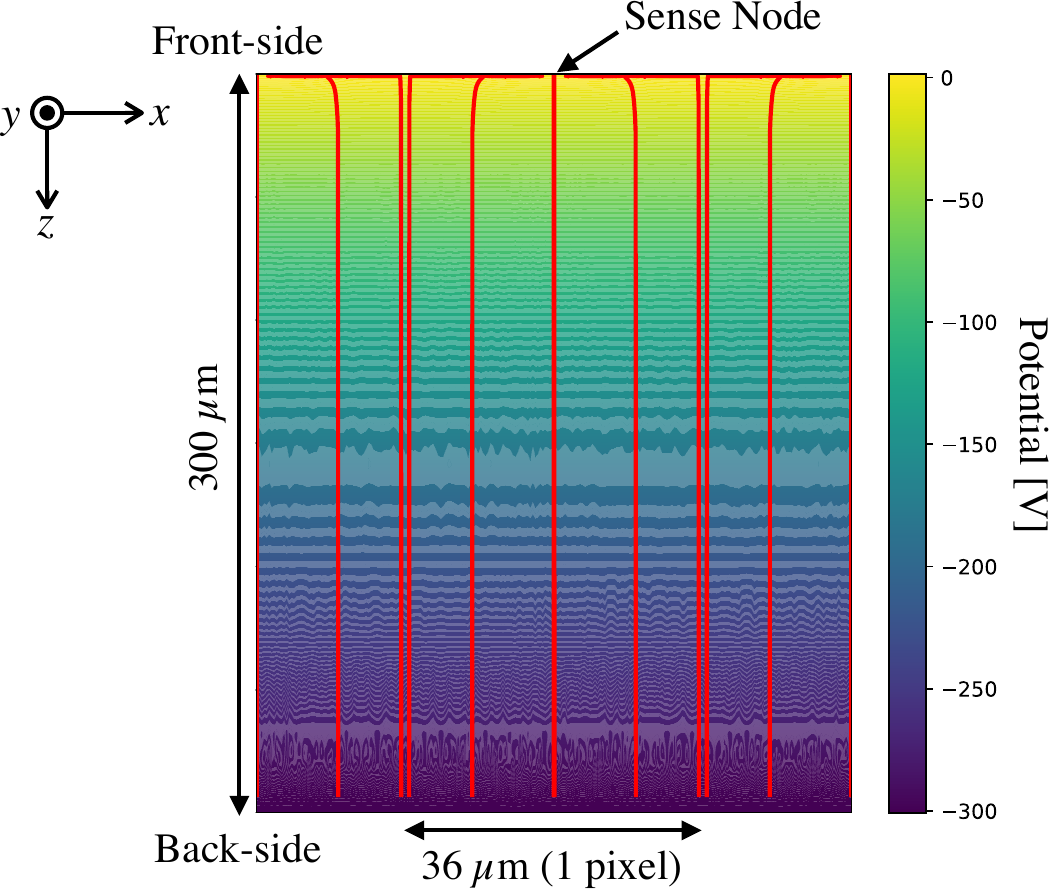}
    \caption{Electrostatic potential map (color map) and electric field lines (red lines) in the sensor layer.}
    \label{fig:06}
\end{figure}

In this method, it is necessary to determine the electron transport path within the depletion region. To determine the transport path, the electric field distribution in the depletion region must be obtained. Therefore, the electric field distribution was calculated using the device simulator HyENEXSS \cite{HyENEXSS}. The calculated electric field distribution is shown in Fig.~\ref{fig:06}.
Since the electric field lines (red lines) are oriented almost entirely along the $z$-direction regardless of the $x$-position, the transport path of electrons generated near the back-side can be assumed to be a straight trajectory from the generation point to the sensor surface along the $z$-direction. Therefore, electron transport in the bulk depletion region is approximated as one-dimensional transport along the $z$-axis. After reaching the sensor surface, the generated charge is collected at the sense node by weak lateral electric field. Therefore, charge loss during the lateral transport near the sensor surface, which does not contribute to the difference between the FI and BI configurations, was neglected, and only charge loss during the z-directional transport in the bulk depletion region was considered.

\begin{figure}[tbh]
    \centering
    \includegraphics[width=0.5\linewidth]{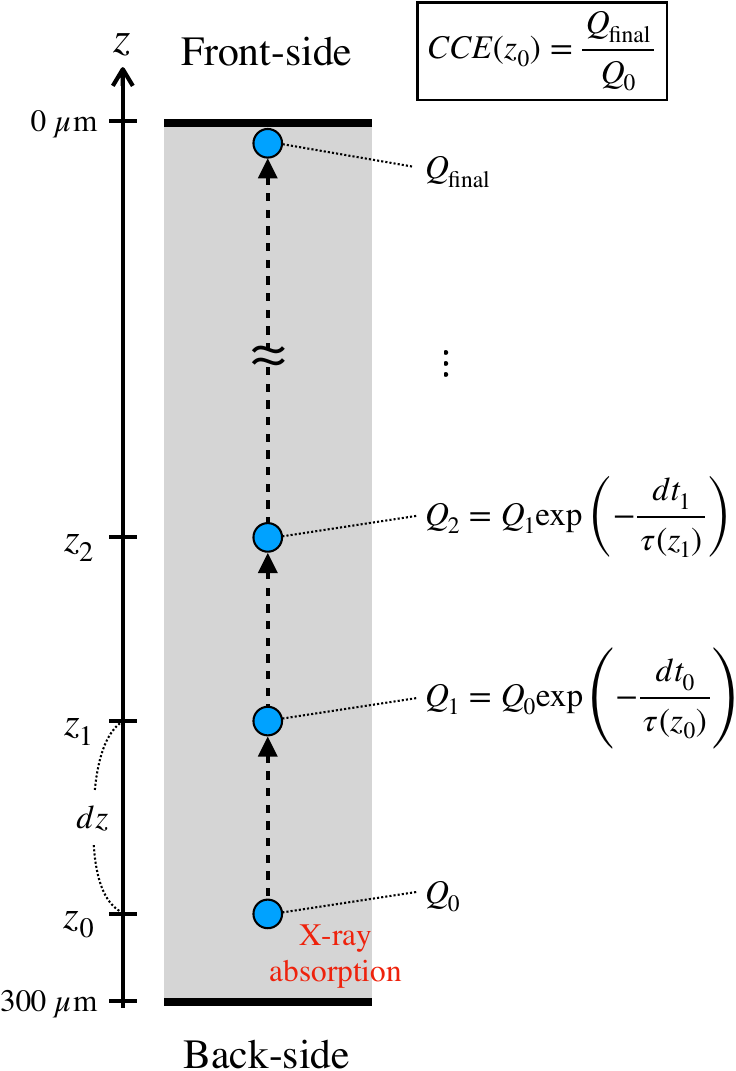}
    \caption{Schematic illustration of the calculation method for the charge collection efficiency (CCE). A point charge generated at an initial position $z_0$ drifts toward the front-side of the sensor while losing charge due to trapping.}
    \label{fig:07}
\end{figure}

As illustrated in Fig.~\ref{fig:07}, a point charge is assumed to move from an initial position $z_0$ by a step size $dz$. Using the $z$-component of the electric field profile at the pixel center, $E(z)$, and the electric-field-dependent electron drift velocity \cite{Caughey},
\begin{equation}
    v(E)=\frac{v_\mathrm{satu}}{\sqrt{1+\left(E_0/E\right)^2}},
\end{equation}
the drift time required to travel the distance $dz$ is given by
\begin{equation}
    dt=\frac{dz}{v(E)}.
\end{equation}
Here, $v_\mathrm{satu}$ is the electron saturation drift velocity and $E_0$ is the critical electric field. The values $v_\mathrm{satu}=1.14\times10^7~\mathrm{cm/s}$ and $E_0=7000~\mathrm{V/cm}$ were used in this study~\cite{Baliga,Sze}.
The electron lifetime at each position is expressed as
\begin{equation}
    \tau(z)=\frac{1}{\sigma_\mathrm{t}v_\mathrm{th}N_\mathrm{t}(z)},
\end{equation}
where $N_t(z)$ is the trap density obtained from the I-V measurement. The remaining charge after drifting by $dz$ is updated iteratively as follows. The charge amount $Q_{i+1}$ after the $dz$ drift at the $(i+1)$-th step is calculated from the charge amount $Q_i$ at the $i$-th step according to
\begin{equation}
    Q_{i+1}=Q_i \exp\left(-\frac{dt}{\tau}\right).
\end{equation}
This calculation is repeated from the initial position $z_0$ until the point charge reaches the sensor surface. The final collected charge, $Q_{\mathrm{final}}$, normalized by the initial charge $Q_0$, is defined as the charge collection efficiency at the initial position $z_0$,
\begin{equation}
    CCE(z_0)=\frac{Q_{\mathrm{final}}}{Q_0}.
\end{equation}
By performing this calculation for all initial positions from the back-side to the front-side of the sensor, the depth-dependent charge collection efficiency, $CCE(z)$, was obtained. A step size of $dz=1~\mu\mathrm{m}$ was used in the calculation. Reducing the step size further changed the calculated CCE by less than $1\times10^{-5}\%$.

\begin{figure}[tbh]
    \centering
    \includegraphics[width=0.55\linewidth]{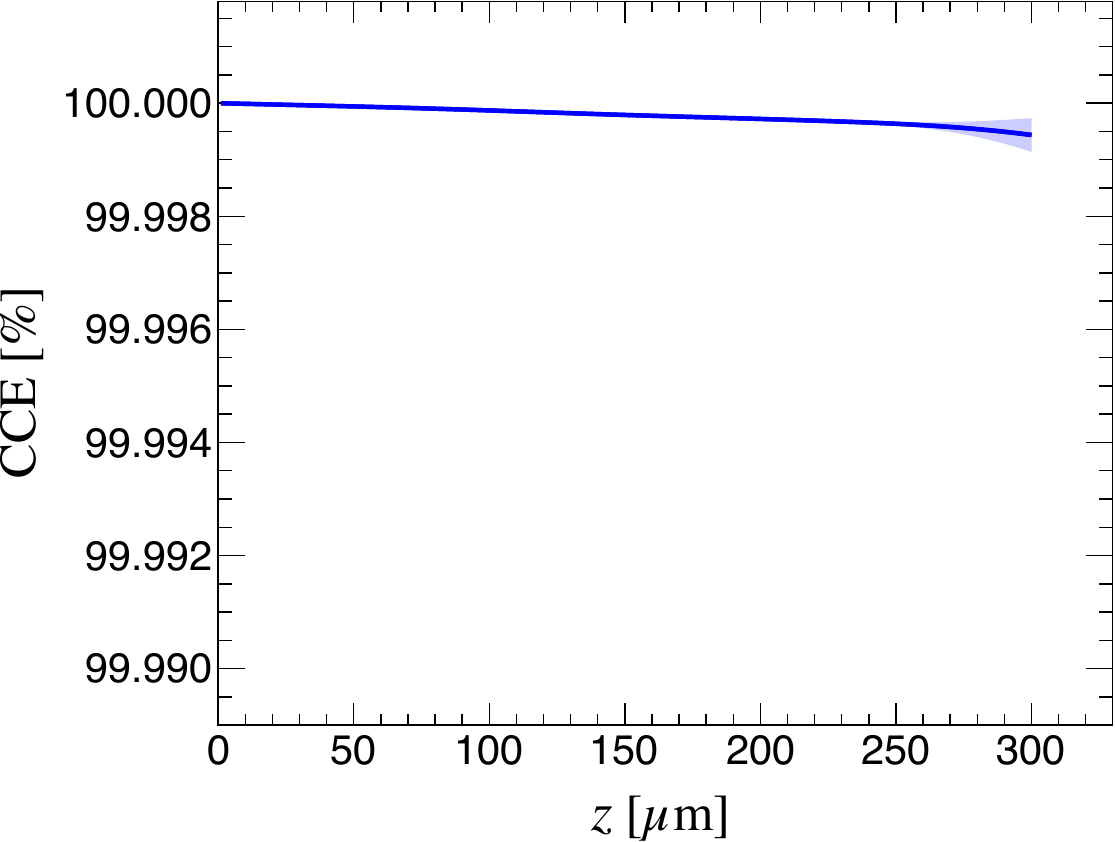}
    \caption{Depth profile of the charge collection efficiency calculated from the trap density distribution. The shaded region represents the uncertainty propagated from the measured trap density.}
    \label{fig:08}
\end{figure}
The calculated depth-dependent CCE is shown in Fig.~\ref{fig:08}. Even near the back-side, the CCE remained higher than approximately 99.9995\%, indicating that charge loss due to the traps near the back-side is negligible. Therefore, charge generated near the back-side is transported to the front-side with virtually no loss. Although the back-side process introduces a region with high trap density near the back-side, the resulting charge loss is negligible and cannot account for the observed peak shift.

\subsection{Effect of charge cloud spreading}
The electric field distribution shown in Fig.~\ref{fig:06} is also essential for evaluating the spreading of the charge cloud generated by X-ray absorption during its drift through the depletion region. This is because the diffusion width of the charge cloud is proportional to the square root of the drift time, which depends on the carrier drift velocity determined by the electric field strength. In addition, the direction of the electric field lines is important because a lateral electric field toward the sense node suppresses charge spreading. As shown in Fig.~\ref{fig:06}, the lateral electric field toward the sense node is negligible over most of the sensor thickness. Therefore, during this drift from the back-side to the front-side, the charge cloud is assumed to spread with an approximately Gaussian distribution owing primarily to thermal diffusion and Coulomb repulsion.
Based on this drift process, the charge cloud size after transport was calculated using the mean photoelectric absorption depth of X-rays \cite{Henke}, the drift time obtained from the electric field distribution, and the effects of both thermal diffusion and Coulomb repulsion. The resulting charge cloud diameter is given by 
\begin{equation}
	\frac{\partial\sigma^2}{\partial t}=2D+\frac{\mu_\mathrm{e}Ne}{12\pi^{3/2}\varepsilon\sigma},
	\label{eq:diff}
\end{equation}
where $t$ is the drift time, $D$ is the electron diffusion coefficient, $\mu_\mathrm{e}$ is the electron mobility, $N$ is the number of generated electrons, $e$ is the elementary charge, and $\varepsilon$ is the permittivity of Si. The values used in the calculation were $D=40~\mathrm{cm^2/s}$, $\mu_\mathrm{e}=1.8\times10^3~\mathrm{cm^2/(V\,s)}$, $e=1.60\times10^{-19}~\mathrm{C}$, and $\varepsilon=1.05\times10^{-12}~\mathrm{F/cm}$ \cite{Benoit}.
\begin{figure}[tbh]
    \centering
    \includegraphics[width=0.5\linewidth]{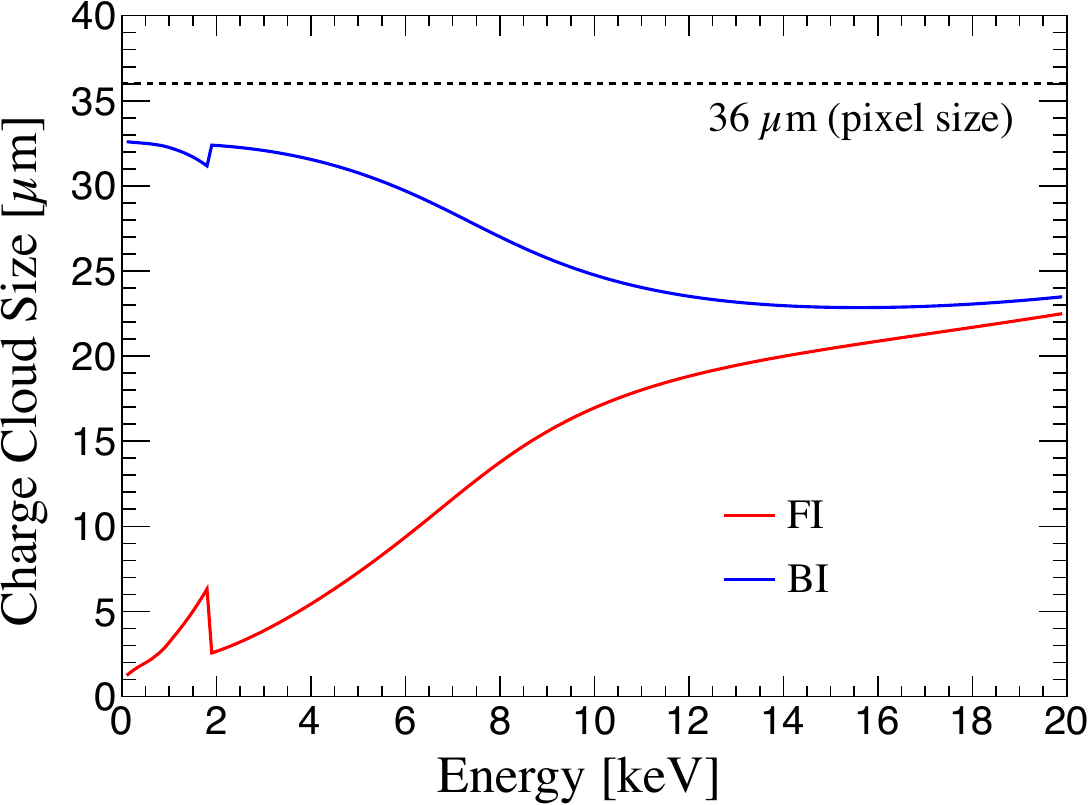}
    \caption{Charge cloud size at the sense node for each X-ray energy in the FI and BI configurations. The diameter corresponds to 6$\sigma$, where $\sigma$ is defined in Eq.~\ref{eq:diff}. Red and blue denote the FI and BI results, respectively, and the black dashed line indicates the pixel size $36~\mathrm{\mu m}$.}
    \label{fig:09}
\end{figure}
Figure~\ref{fig:09} shows the calculated charge cloud diameter at the sensor surface for both FI and BI as a function of the incident X-ray energy. In the BI configuration, the charge cloud diameter is larger than that in the FI configuration at lower X-ray energies and approaches the pixel size of $36~\mathrm{\mu m}$. For example, at the lowest evaluated X-ray energy of 5.89~keV, the calculated charge cloud diameter is approximately $10~\mathrm{\mu m}$ for FI, whereas it reaches approximately $30~\mathrm{\mu m}$ for BI. Consequently, in the BI configuration, part of the generated charge is more likely to spread into neighboring pixels before being collected at the sensor surface. When the charge transferred to the neighboring pixels is below the split threshold, the event is still classified as a single-pixel event. As a result, the pulse height recorded for the center pixel is reduced by the leaked charge, causing the peak position of single-pixel events to shift toward lower energies. This mechanism is considered to be a possible origin of the low-energy peak shift observed for the BI device.

\begin{figure}[tbh]
    \centering
    \includegraphics[width=0.5\linewidth]{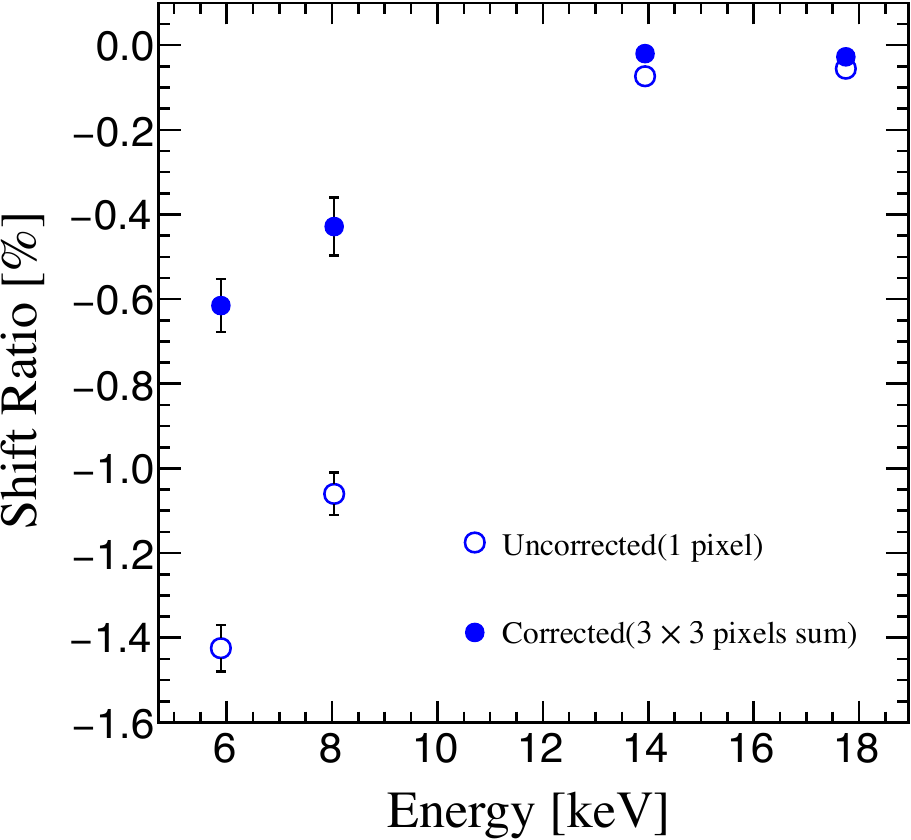}
    \caption{Dependence of the shift ratio on X-ray energy before and after the 3$\times$3 pixel-sum correction. Open blue circles indicate the results before the correction, and filled blue circles indicate the results after the 3$\times$3 pixel-sum correction.}
    \label{fig:10}
\end{figure}

To verify this interpretation, the peak shift was recalculated using the sum of the pulse heights of the event-center pixel and the surrounding 3$\times$3 pixels. This approach includes charge signals below the split threshold; however, it also introduces additional noise from surrounding pixels that do not contain signal charge, potentially degrading the signal-to-noise ratio (S/N). The recalculated results are shown by the blue plots in Fig.~\ref{fig:10}.
As can be seen from Fig.~\ref{fig:10}, the peak shift is reduced over the entire energy range when the summed pulse height of the 3$\times$3 pixels is used. In particular, at the lowest measured X-ray energy of 5.89~keV, the peak shift is reduced from approximately 1.4\% to approximately 0.6\%. These results indicate that accurate determination of the peak position for low-energy X-rays in the BI configuration requires consideration of charge cloud spreading based on the electric field distribution. Furthermore, when the charge cloud diameter becomes comparable to the pixel size, data analysis that accounts for charge sharing with neighboring pixels effectively suppresses the peak shift observed in the BI spectra.

\section{Conclusion}
We compared the spectra obtained with the FI and BI configurations to understand the X-ray response of XRPIX in the BI. The results showed that the X-ray peak positions in the BI shifted toward lower energies, with a larger shift for lower-energy X-rays. A peak shift of approximately 1.4\% was observed at 5.89~keV. The origin of this peak shift was investigated, and the contribution of charge loss due to back-side defects was found to be negligible. The results suggest that the peak shift is mainly caused by the reduction in collected charge due to charge sharing associated with charge cloud spreading. Furthermore, by considering charge cloud spreading in the data analysis, the peak shift ratio was reduced to approximately 0.6\% at 5.89~keV. The findings of this study are expected to be useful for high-precision energy calibration and the construction of detector response models for future X-ray astronomical missions using BI-type XRPIX.





\section*{Acknowledgements}
We acknowledge the valuable advice and great work of the personnel of LAPIS Semiconductor Co., Ltd. This research was supported in part by MEXT/JSPS KAKENHI Grant-in-Aid for 22H01269 and 23K22540.

\bibliographystyle{elsarticle-num} 
\bibliography{reference}





\end{document}